\documentclass[preprint,showpacs,preprintnumbers,amsmath,amssymb,prl]{revtex4}

\usepackage{graphicx}
\usepackage{bm}

\usepackage{amsmath,amssymb}

\begin{document}

\preprint{}

\title{The Effect of Disease-induced Mortality on Structural Network Properties}

\author{Lazaros K. Gallos}
\author{Nina H. Fefferman}
\affiliation{
DIMACS, Rutgers University, Piscataway NJ 08854, USA, and Department of Ecology, Evolution, \& Natural Resources, Rutgers University, New Brunswick
NJ 08901, USA.}

\date{}

\begin{abstract}
As the understanding of the importance of social contact networks in the spread of infectious diseases has increased,
so has the interest in understanding the feedback process of the disease altering the social network. While many studies
have explored the influence of individual epidemiological parameters and/or underlying network topologies on the resulting
disease dynamics, we here provide a systematic overview of the interactions between these two influences on population-level
disease outcomes. We show that the sensitivity of the population-level disease outcomes to the combination of
epidemiological parameters that describe the disease are critically dependent on the topological structure of the population's
contact network. We introduce a new metric for
assessing disease-driven structural damage to a network as a population-level outcome. Lastly, we discuss how the expected
individual-level disease burden is influenced by the complete suite of epidemiological characteristics for the circulating
disease and the ongoing process of network compromise. Our results have broad implications for prediction and mitigation of
outbreaks in both natural and human populations.
\end{abstract}

\maketitle

\section{Introduction}
Recent advances in our understanding of complex social structures have led to a re-evaluation of epidemiological processes
taking place on these structures \cite{Brockmann,Saramaki,Verdasca,Bansal2007,Balcan,Goltsev,Black,Keeling}. Almost all infection models have been shown to behave differently on complex networks
compared to simple lattice structures or to fully mixed models \cite{Pastor2001,Eubank2010}, reflecting different potential types of contact patterns
among various populations \cite{Eubank2004,Sah,Jones,Gallos2012}. Typical epidemiological models \cite{Anderson} include variations of the basic SI, SIR, and SIS models,
where susceptible (S) individuals can become infected (I) upon encounter with other infected individuals and eventually either
recover with immunity (R-state where they cannot be re-infected) or without immunity (i.e. return to the susceptible state).
Each of these models is appropriate to describe varying conditions of spreading.

Thus far, however, no network-based analysis has considered cases in which the disease both generates a limited-duration
immunity that eventually lapses back into susceptibility
(due either to genetic drift, including antigenic drift \cite{Smith} of the pathogen
or to loss of T-cell memory \cite{Lofgren}, while also carrying a non-trivial risk of disease-induced mortality \cite{Franke}.
While both of these examples focus on influenza virus \cite{Shaman,Bolton},
many pathogens exhibit this pattern of generation of immunity that later wanes.

Explicit study of such a case may, in fact, be of particular practical importance since both disease-related death and temporary
immunity will interrupt successful disease transmission over the remaining network. The dynamics between permanent removal
(i.e. death) and the temporary removal (i.e. immunity) may drive the emergence of very different global patterns in disease outcomes. 

As has been well-studied \cite{Pastor2014,Bansal2010,Handcock}, the interplay between the network structure and the dynamic spreading process strongly influences
the outcomes of an epidemic. In lattice structures, all nodes have a similar importance to disease risk and spread because of the
spatial invariance and the homogeneous character of the system, so that these effects do not exist. In contrast, in complex network
the structure is dominated by the existence of well-connected hubs. These nodes are included in the majority of all possible paths,
so that a disease can easily reach them. Even nodes that are not well-connected can become very significant in spreading if they happen
to be in an appropriate location \cite{Kitsak}. Of course, for pathogens that carries a non-trivial mortality risk, frequent infection of a node
in any network will eventually lead to its removal. This removal greatly impacts the structural features of the remaining network.
The topology becomes more hostile to spreading and areas that were easily connected through the hubs can now become protected by the
disease simply by isolation. This isolation, though, can have detrimental effects on proper communication in the network. In short,
the interplay between the exposure of nodes to infection and their asymmetrical impact of removal on both topology and dynamics,
creates a complex cycle with unusual epidemiological properties.

Disease-induced mortality itself may be especially important to consider when non-disease-related processes of network function may be
drastically diminished by disease-induced structural compromise, even though exactly such disconnection acts to diminish the probability
of transmission of future infection for the remaining nodes (by decreasing effective population density) \cite{Ferrari}.  In these cases, some
standard measures of structural integrity of the population may seem uncompromised (e.g. largest remaining connected component) \cite{Cohen2010},
even though function can be reduced to the point of failure (e.g. increased average minimal path lengths for communication between
individuals in the population) \cite{Lopez}. To study these types of functional effects, we introduce a new measure, the stability index,
which can take into account partial structure compromise as a result of an infectious epidemic with an associated mortality risk.
This consideration may be useful in fields such as conservation biology or communication networks.

\section{Methods}
\subsection*{The SIRDS model.}

\begin{figure}[h]
\centerline{\resizebox{9cm}{!} { \includegraphics{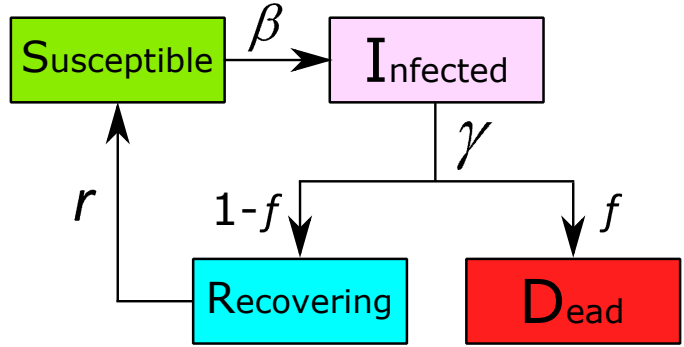}}}
\caption{{\bf Schematic of the SIRDS process.}}
\label{fig1}
\end{figure}

We model the spreading of a potentially fatal epidemic disease in a population. Recovery from the disease provides short-term
immunity \cite{Brauer}. The disease originates in a randomly selected node (then automatically designated in the infected class, I)
in an otherwise fully susceptible population, S. In every time-step, all the susceptible neighbors of the infected nodes become
infected with probability $\beta$ per infected neighbor. After attempting to infect their neighbors, the infected nodes leave
the I class at a recovery rate $\gamma$. These nodes either die with probability $f$, so that they fall in the D (Deceased), or recover
into the R (Recovered) class with probability $1-f$. Nodes in recovery are immune-protected, losing that protection at a rate $r$,
the rate of loss of protection.
The number of surviving nodes in this model is a function of time, $N(t)$, with the initial population being $N(0)=N$ nodes.
We describe this SIRDS process in Fig.~\ref{fig1}. The equations that describe this model are:

\begin{align}\label{eq1} 
\dot{S} = -\beta S I + rR \\
\dot{I} = \beta SI - \gamma I \\
\dot{D} = \gamma f I \\
\dot{R} = (1-f)\gamma I - rR \\
N(t) = S(t)+I(t)+R(t) = 1 - D(t)
\end{align}

The mortality probability, $f$, determines what percentage of the infected population will leave the system. The two limiting cases
of the SIRDS model correspond to a simple SIR model, when $f=1$, and to an SIRS model, when $f=0$. There are no ongoing demographic
processes, such as natural births or natural deaths in the model, so death results only as an outcome of the modeled disease.
We fix the rate $\gamma$ to $\gamma=1$, which also fixes the time-scale of the system {so that our time unit throughout the
simulations is equal to the recovery time}. In the following, we study the effect of the remaining parameters in the system, infection probability, $\beta$, mortality
probability, $f$, and protection loss rate, $r$.

The main quantity of interest in our current study is the mass of dead individuals, $D$, and the conditions which determine its
asymptotic, $D(\infty)$, and finite-time value, $D(t)$. We will use this quantity to further estimate the probability that the
population will retain its long-range connectivity. This is a measure of the viability of the population and its ability to
survive against spreading of fatal infections.

\subsection*{Network structure.}

We study the SIRDS process on two typical structures for the initial population connectivity: a square two-dimensional lattice
and a scale-free network. We further split the scale-free network examples into two cases: a) random scale-free networks,
which have been used to describe large-scale human populations \cite{Chakrabarti}, and b) self-organized networks, which may be more reflective
of emergent structures in many natural populations \cite{Fefferman}. In all cases, we study small networks of size $N=200$, compromising between
an accurate order of magnitude for many natural populations of concern for ongoing persistence and sufficiency of size to enable
meaningful computational observations. We have also verified that the results are not significantly influenced when we increased the
size to $N=1000$ (see e.g. Fig.~\ref{fig2}).

The random scale-free networks are created by the configuration model \cite{Molloy}, with a power-law degree distribution of exponent $\lambda=2.5$,
i.e. $P(k)\sim k^{-2.5}$. In this structure most nodes have a low degree but the hubs are quite strong, with each hub connected
to roughly 5-30\% of the network.

The self-organized networks describe a different social organization, where a few nodes act as super-hubs and are connected to
almost every other node in the system \cite{Fefferman}. This simulates a strongly hierarchical society, where a few `alpha' animals dominate
over the entire group (e.g. grooming behaviors in primates, \cite{Madden}). We chose the model parameters so that we remain
consistent with previous research on such structures \cite{Fefferman}. This network is built as follows: All nodes start with an initial degree $k=5$, with 5 randomly selected nodes as neighbors.
We assume that the nodes evolve their connections, with the goal of connecting to the most-connected nodes. Therefore,
at each step all nodes remove their two neighbors with the smallest degree and create new links towards two randomly selected nodes.
In this way, the nodes preferentially attach themselves to the largest hubs, which become progressively larger until at the end they
are connected to almost the entire network. We use this final form of the network as the static representation of connectivity,
and we apply the SIRDS process on this structure.

\subsection*{Survival probability.}

We characterize the survival of nodes through the survival probability, $\Phi$. Here, $\Phi$ is defined as the probability that a node
that is alive at the present state of the spreading process will remain alive until the disease has died out. We consider this property
to be a function of either time, $\Phi(t)$, or of the percentage of removed nodes, $\Phi(p)$. These two parameters describe the state
of the network, from a different perspective. The first parameter is the number of steps, $t$, since the beginning of infection,
and this measure can be useful to determine how much time is available to intervene, independently of the current network damage.
The second parameter is the fraction $p$ of the initial network that has been removed due to the disease, independently of the time
required to reach this level of damage. In practice, we quantify the evolution of the survival probability by the point
$t_{0.9}$ or $p_{0.9}$ when $\Phi$ first becomes equal or larger to 90\%, $\Phi>0.9$, i.e. after the point where more than 90\% of
the remaining nodes will eventually survive.

\subsection*{Robustness.}
The resilience of a network with regards to node removal is typically measured through the size of the largest cluster remaining,
$S_{\rm max}$, compared to the size of the largest cluster before removing any nodes \cite{Cohen2000}. The connectedness of a cluster does not
guarantee, though, the efficient operation of the network. For example, the fact that one node can still reach another node may not
be as important as the fact that the path length between two nodes has increased so much that it is no longer meaningful to consider
the two nodes reachable from each other. This behavior has been described by the concept of limited path percolation (LPP) \cite{Schneider}.
In LPP, two nodes which are originally at distance $\ell_{ij}$ from each other are considered to be connected after the attack only if their
new distance is smaller than $a\ell_{ij}$. The parameter $a$ indicates our tolerance of the communication distance. A value of $a=1$
requires that the original distances remain intact for the nodes to be considered connected, while $a=\infty$ implies that the distances
are no longer important and this case coincides with the typical identification of the largest connected component.

Here, we suggest a combination of two separate ideas to describe the efficiency of the remaining connected cluster.
First, instead of using the fraction of nodes in the largest cluster, we calculate the area under the dynamic calculation of this fraction
$\int_0^{p_{\rm max}} S_{\rm max}(a) dp$.
This method has been introduced to estimate the efficiency of an attack strategy, independently of the final value $S_{\rm max}$ \cite{Schneider}.
A key point in our method is that the evolution of the node removal is not described by time, for example by the number of steps.
Instead, we use the fraction of removed/dead nodes, $p$, which is a direct result of the disease-induced mortality process.
In this way we can directly compare different processes on different networks when the same number of nodes has been removed in each case.
Another quantity that characterizes the spreading process is the maximum number of nodes that have been removed due to infection, $p_{\rm max}$,
until there is no infection in the system. Notice that the largest cluster does not necessarily vanish at $p_{\rm max}$, and this is what
separates this index from the unique robustness measure, introduced in \cite{Schneider}. That definition cannot be applied if the removal process terminates
before the largest cluster vanishes. Additionally, the maximum possible value for the integral is not necessarily 0.5, and proper normalization
needs to take this fact into account. Therefore, the quantity of interest is the area under the curve of $S_{\rm max}$ from $p=0$ to $p_{\rm max}$.
We further normalize this quantity by the area under the case of minimum possible damage, $1-p$, where the only nodes that leave the spanning cluster
are those that have been physically removed. As a result, we define a stability index, $B(a)$, as

\begin{equation}
B(a) = \frac{\int_0^{p_{\rm max}} S_{\rm max}(a) dp}{\int_0^{p_{\rm max}} (1-p) dp}
\label{eq2}
\end{equation}

The limits of this expression are a) $B=1$ when the least possible damage has been done and the only nodes missing from the largest cluster
are those that have been removed by the disease, and b) $B=0$ when the largest cluster vanishes immediately after removing a few nodes.
Therefore, the stability index can characterize the extent of damage in the remaining largest cluster, independently of its final size.
The stability index coincides with the unique robustness measure \cite{Schneider} when $p_{\rm max}=1$ and $a=\infty$.
The second key idea that we use in this definition comes from the Limited Path Percolation method \cite{Lopez}, through the parameter $a$.
This parameter allows us to characterize the same cluster under varying requirements for functionality. The typical case where connectivity
between any two nodes is enough for network function, independently of how long the distance between these nodes has become,
is expressed by the value of $a=\infty$. We use this case to normalize the results. The ratio of $B(a)/B(\infty)$ then, indicates our
possible error when we decide about the survival of a group strictly from the existence of a spanning cluster. An important feature
of this error is that it includes information from the entire process and not only from the final state. For example, damage at earlier
stages leads to a smaller $B$ value.

\subsection*{Computational experiments}

We apply the SIRDS model for three different rates of loss of protection, $r$: $r=0.05$, 0.20, and 0.50. Each of these values of $r$ indicates a different
duration of immunity for an already infected node. In the first case, a recovering node remains immune for 20 time steps, presenting a natural
obstacle for spreading over a significant amount of time. In the second case, immunity lasts for an intermediate interval of 5 steps,
while in the latter case the node becomes susceptible after only 2 steps. We independently vary the probabilities $f$ and $\beta$ from 0.05 to 1,
in steps of 0.05. For each case we average over 20 different realizations of the structure. In each realization every node serves as the infection
origin 5 times, so that each point has been averaged over a total of 100,000 simulations of the epidemic process. For each case we record the
fraction of the population, $D$, that died because of the disease and the duration of the epidemics, $T$, defined as the time from the initial
infection until when there is no infected individual. All simulations were run until the infection died out, independently of the number of steps
required to reach this stage or the number of infected/diseased nodes. The code for the simulation of the SIRDS is freely available
at \url{http://www.rci.rutgers.edu/~feffermn/code.php}.

Parameter ranges were chosen to explore a sufficient diversity of epidemiological characteristics to
demonstrate how different diseases may produce substantially different results, and we recommend that specific
analyses for particular diseases make use of rates tailored to the specific population/network of interest.

\subsection*{Notation}

In Table \ref{table1} we summarize the notation that we use throughout the paper:

\begin{table}
\begin{tabular}{|c|l|}
\hline
Symbol & Definition \\
\hline
$t$ & Number of time steps in the simulation \\
$S(t)$ & Number of susceptible individuals at time $t$\\ 
$I(t)$ & Number of infected individuals at time $t$\\
$R(t)$ & Number of recovering individuals at time $t$\\
$D(t)$ & Number of dead individuals at time $t$\\
$N(t)$ & Number of surviving individuals at time $t$ \\
$\beta$ & Infection probability \\
$\gamma$ & Recovery rate \\
$f$ & Probability of death for an infected individual \\
$r$ & Loss of protection rate \\
$T$ & Maximum epidemic duration (time until $I(t)=0$) \\
$p$ & Fraction of dead individuals $=D(t)/N(0)$ \\
$p_{\rm max}$ & Maximum value of $p$, at $t=T$ \\
$\lambda$ & Degree exponent for the scale-free networks \\
$\Phi(t)$ & Probability for a node to survive until $T$, given that it is alive at time $t$ \\
$\Phi(p)$ & Probability for a node to survive until $T$, given that $p$ nodes have been removed \\
$t_{0.9}$ & Number of time steps when the survival prob. first becomes $\Phi(t)\geq0.9$ \\ 
$p_{0.9}$ & Fraction of removed nodes when the survival prob. first becomes $\Phi(p)\geq0.9$ \\ 
$S_{\rm max}$ & Fraction of surviving nodes that form the largest remaining cluster \\
$\ell_{ij}$ & Shortest path distance between nodes i and j in the original network \\
$a$ & Nodes are considered disconnected if their distance becomes $>a\ell_{ij}$ \\
$B(a)$ & Stability index, defined in Eq.~\ref{eq2} \\
\hline
\end{tabular}
\caption{ \label{table1} Definitions of the main parameters and properties used in the paper.
}
\end{table}

\section{Results}

\subsection*{Fraction of dead population}

\begin{figure}[h]
\centerline{\resizebox{15cm}{!} { \includegraphics{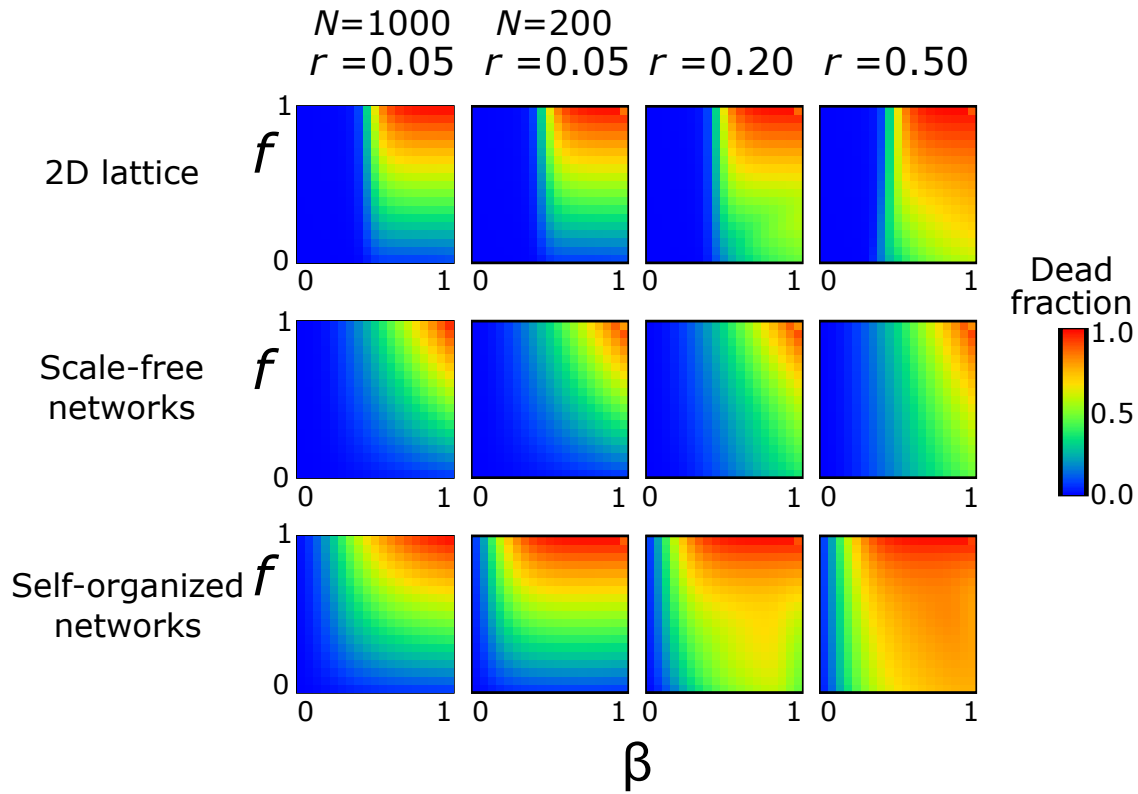}}}
\caption{{\bf Fraction of the population that died because of the disease, as a function of the infection probability and the death probability.}
From top to bottom: two-dimensional lattices, random scale-free networks ($\lambda=2.5$), and self-organized networks.
The x-axis corresponds to the infection probability, $\beta$, and the y-axis to the death probability, $f$, of an infected node.
The columns correspond to the protection loss rate, $r$, of a node (left to right): $r=0.05$ ($N=1000$), $r=0.05$ ($N=200$), $r=0.20$, and $r=0.50$.
The threshold point for a given combination of parameters is defined as the point where the diseased fraction becomes larger than zero,
independently of the infected mass. In the plots, the threshold values for each case can be found at the point where the blue area turns into green.}
\label{fig2}
\end{figure}

For a two-dimensional lattice the picture is very similar to what we would expect from a standard SIR model (Fig.~\ref{fig2}, top row).
In the SIR model (which corresponds to the SIRDS model when $f=1$), there is a sharp transition as we increase the infection probability,
$\beta$, from a `safe' population with almost no mortality to nearly complete annihilation at $\beta>0.5$. A similar pattern is observed here
in the results for the lattice.
The mortality probability, $f$, has little influence, as long as it has a value that
is not close to 0, e.g. $f>0.1$. As we increase the protection loss rate for the same infection and mortality rates,
the influence of $f$ becomes weaker and a larger part of the
population dies: in a faster recovery the nodes spend more time in the susceptible state where they can be infected rather than in
the recovering state, where they are immune.

The picture is quite different in random scale-free networks (Fig.~\ref{fig2}, middle row). The region of total annihilation is now
restricted to high values of both $f$ and $\beta$. Here, we consider the threshold point for epidemics to be the combination of parameters
where the diseased fraction becomes larger than zero, independently of the infected mass. Even though the threshold for an epidemic outbreak remains close to $\beta=0.5$,
the inflicted damage is considerably smaller than in a lattice. The hubs are connected to a significant fraction of the network,
while the majority of the nodes have very few connections. These conditions result in efficient protection of the population.

In contrast, the hubs in self-organized networks are much stronger and are connected to almost all other nodes (Fig.~\ref{fig2}, bottom row).
This makes the network more vulnerable, even for low mortality probabilities $f$. The percolation threshold for $\beta$ is considerably smaller
than in lattices and scale-free networks. The threshold value is now close to $\beta=0.1$, which indicates that it is much easier for
epidemics to occur because of the extremely centralized nature of the network. Moreover, similar results are obtained even when $f$ is very low,
in particular for large protection loss rates, $r$. Indicatively, when $\beta=0.3$ and $f=0.3$ at $r=0.50$ the infection leaves almost 55\% of the
population dead, while the corresponding fraction in scale-free networks is around 6\% and in lattices it is 0.3\%. These differences point out
the different structural character of each system and its influence on mortality due to epidemic spreading.

The influence of the protection loss rate $r$ on the results is mainly quantitative. The behavior of the dead fraction does not change a lot
as we increase the rate of protection loss for the same structure, and the general features that we find in the plots for small $r$ also apply to those of large $r$. In the
following sections, we also find that $r$ mainly influences the numerical values of the epidemic duration and the survival probability, but has
otherwise a limited effect.

\subsection*{Duration of outbreak}

\begin{figure}[h]
\centerline{\resizebox{15cm}{!} { \includegraphics{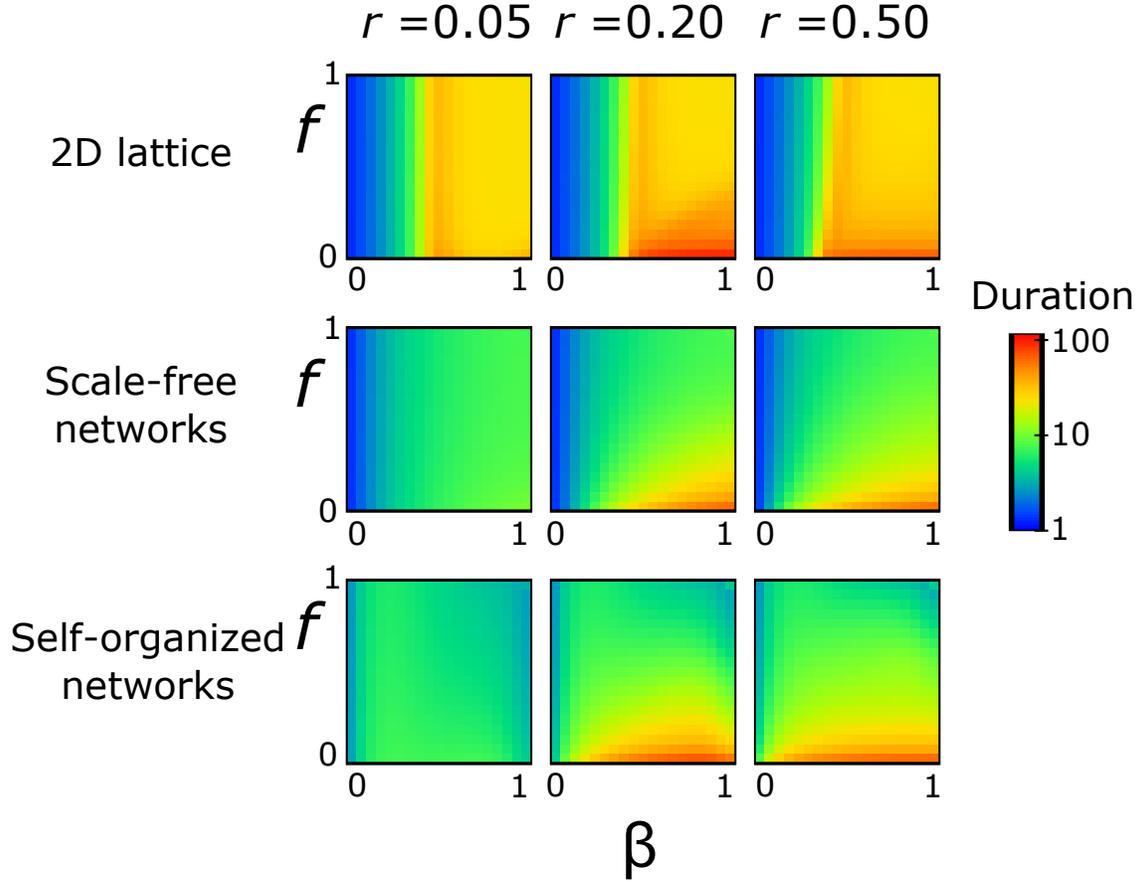}}}
\caption{{\bf Duration of epidemic: time till stochastic die-out.}
From top to bottom: two-dimensional lattices, random scale-free networks ($\lambda=2.5$), and self-organized networks.
The x-axis corresponds to the infection probability, $\beta$, and the y-axis to the death probability, $f$, of an infected node.
The columns correspond to the protection loss rate, $r$, of a node (left to right): $r=0.05$, $r=0.20$, and $r=0.50$.}
\label{fig3}
\end{figure}

An important feature of the spreading process is the duration of the epidemics. A longer duration leaves a much larger time window for
possible intervention, while a shorter duration may complete the maximum spreading cycle before any action can be taken.
In lattices, the duration is dictated by the value of $\beta$ and is almost independent of $f$, except for large $\beta$ and small $f$
values where we observe somewhat longer durations (Fig.~\ref{fig3}, top row). This is in contrast to scale-free networks (Fig.~\ref{fig3}, middle row),
where three regimes are found: a) low-$\beta$ regime: the disease lasts only for a couple of steps and dies rapidly without causing any damage,
b) intermediate-to-large $\beta$ regime and large $f$ values: the duration of the epidemics in the $N=200$ network is of the order of 10 steps.
Even though the duration is small, the damage is considerable, and c) intermediate-to-large $\beta$ regime and small $f$ values: the epidemics now
can last for more than 100 steps, even though it is not as lethal as the previous case. In self-organized networks (Fig.~\ref{fig3}, bottom row) the
picture is basically the same as in scale-free networks, but now the epidemics may spread extremely fast even at large values of $\beta$. Interestingly,
even though the duration may change by an order of magnitude as we vary the mortality probability, the end result is always a large fraction of the
population dying (from 75-100\%). 

\subsection*{Classification of the outbreak}

Figs.~\ref{fig2} and \ref{fig3} suggest the existence of roughly four regimes for the corresponding probabilities:
a) Low mortality -- Low infectivity, b) Low mortality -- Large infectivity, c) Large mortality -- Low infectivity, and d) Large mortality -- Large infectivity.
Lattices and scale-free networks are quite similar qualitatively in all these regimes, with large damage when both $\beta$ and $f$ are large.
The distinguishing feature of self-organized networks is that the dead fraction can be very large when only one of the two basic parameters, $\beta$ or $f$,
is large, even if the other one remains relatively small.

\begin{figure}[h]
\centerline{\resizebox{15cm}{!} { \includegraphics{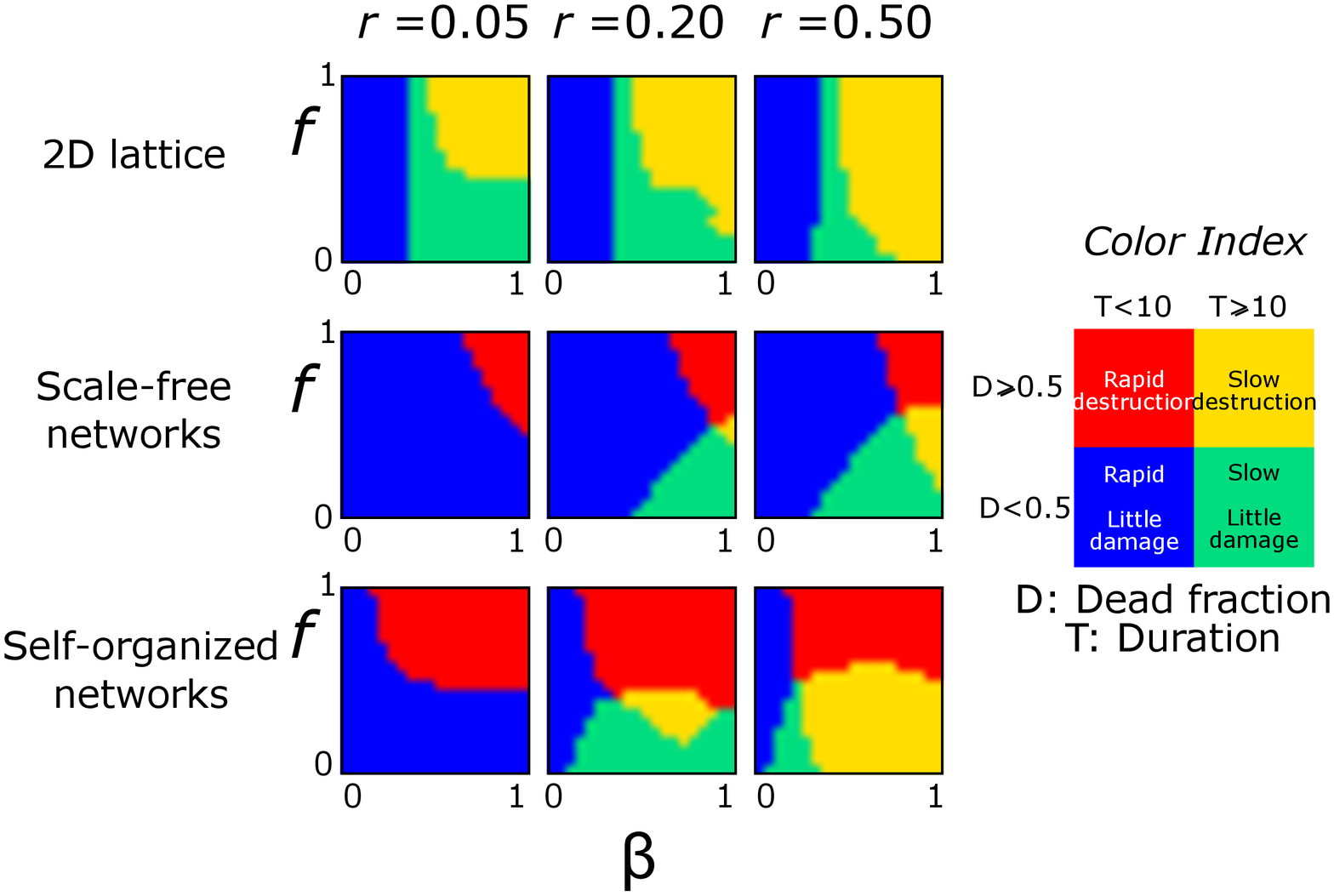}}}
\caption{{\bf Combination of the results on dead population and duration.}
The color in the plots, as explained in the index, indicates whether the infection destroyed more or less than half the population
and whether it lasted more or less than 10 steps. Red and yellow indicate the areas of larger damage.}
\label{fig4}
\end{figure}

In Fig.~\ref{fig4} we combine both the extent of damage and the outbreak duration. For every pair of $\beta$ and $f$ parameters we assign a color depending
on whether more than half of the population died because of the disease and whether the outbreak lasted more or less than 10 steps. In this way, there are
four possible classifications of short/long duration combined with large/small damage. Notice that the duration of the epidemic is only determined
by the number of steps until there is no infected node. A long or short duration is possible even if the end result of a spreading process
is that there are no diseased nodes.

The case of smallest damage is when the dead fraction is low and the disease dies out quickly (blue color in the plots). A trivial result is that this behavior dominates
when $\beta$ is very small, independently of the value of $f$, because a low infection probability drastically limits the infection spread
from the node of initial disease introduction.

In lattices, the process evolves slowly, except when there is little to no node removal. This long duration can lead either to extended damage (yellow areas)
for large $\beta$ and $f$ values or to limited damage (green areas) when $f$ is smaller. Lattices never exhibit short duration/extended damage combination
(red areas) under any combination of parameters, which is the result of the extended spatial distances from the absence of hubs in the structure.

Interestingly, in scale-free networks the dominant area of short duration/little damage extends over much larger values of $\beta$,
as a result of the small damage in general, showing an alternative behavior only under large $\beta$ and $f$ values (Fig.~\ref{fig2}). The case of long duration
with extended damage only emerges in a narrow range of very large infection probabilities and moderate mortality rates in scale free networks. This shows that
for scale-free networks extended damage occurs very rapidly, and only for large values of $\beta$ and $f$. Otherwise, the infection either dies out quickly or
is not able to destroy a significant part of the network.

In self-organized networks, the hubs strongly dominate the structure and create a much smaller-world.  This leads to a generally short duration. For small
protection loss rates, $r=0.05$, the short duration/little damage area dominates the plot, showing extended damage only when mortality rates are extremely high.
For higher protection loss rates, a longer duration was observed for small mortality rates, $f$, which switched from small damage to higher damage as we moved
from $r=0.2$ to $r=0.5$.

\subsection*{The survival probability}

\begin{figure}[h]
\centerline{\resizebox{15cm}{!} { \includegraphics{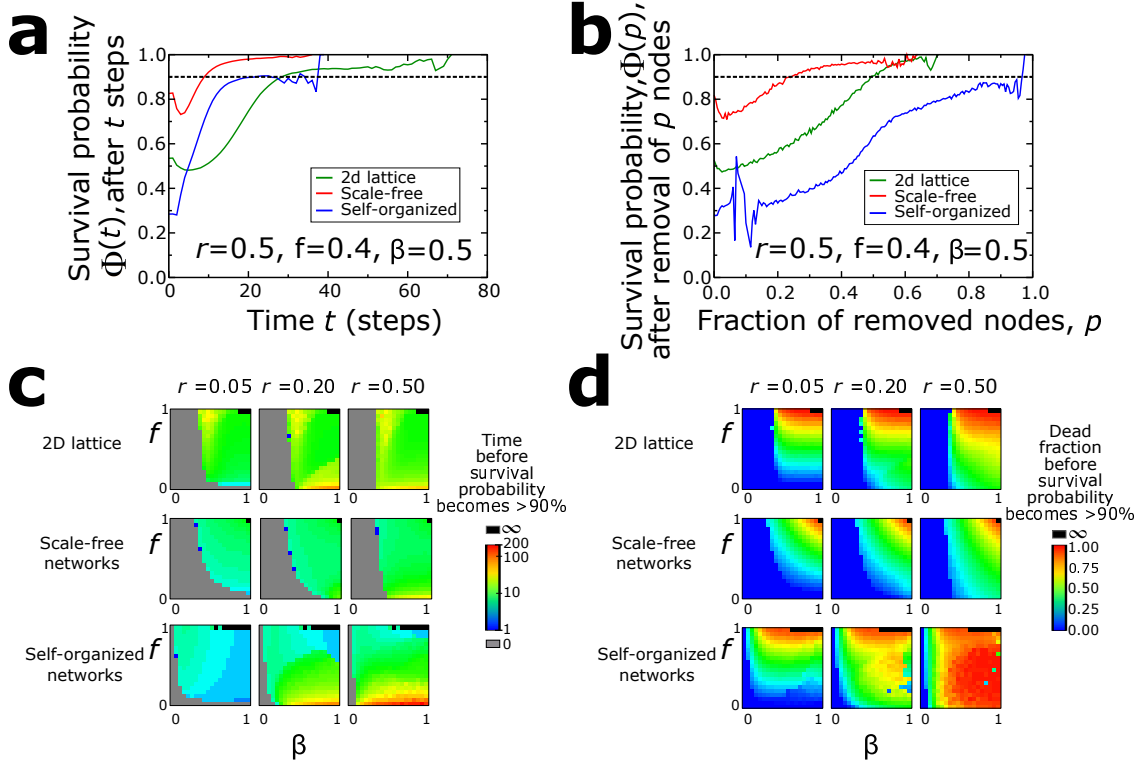}}}
\caption{{\bf Survival probability.} (a) Average survival probability for a node that has survived after $t$ steps as a function of $t$. The lines correspond to the three different topologies,
for a given set of parameter simulations. (b) Average survival probability for a node that has survived after a fraction, $p$, of nodes has died of the disease.
(c) The time $t_{0.9}$ for different parameters $r$, $f$, and $\beta$. (d) The fraction $p_{0.9}$ for different parameters $r$, $f$, and $\beta$. Combination
of the results on dead population and duration.}
\label{fig5}
\end{figure}

The survival probability of a node increases, for the most part monotonically, both with time and $p$. This is the expected behavior: as time passes,
a larger number of nodes die and the network becomes increasingly sparse. Large parts of the network are effectively isolated from the disease,
so that the remaining nodes are less exposed to further infections. In Figs.~\ref{fig5}a and \ref{fig5}b we compare the survival probability $\Phi$
for different topologies. We can see that in lattices, for example, the survival probability $\Phi(t)$ increases at a slower rate than
in (e.g.) self-organized networks. This indicates that if we focus on the duration of the epidemics only, a node in a lattice remains potentially
vulnerable to infections for a longer time. In self-organized networks this process is much more rapid. On the other side, when we consider $\Phi(p)$
as a function of the percentage of removed nodes, it takes a much larger number of removals for a node in self-organized networks to start feeling `safer'.

The results for the $t_{0.9}$ and $p_{0.9}$ points are shown in Figs.~\ref{fig5}c and \ref{fig5}d for all combinations of the model parameters, $r$, $\beta$, and $f$.
It is possible that the epidemic never kills more than 10\% of the population, in which case $t_{0.9}=0$, and there is no time when the probability of
survival is less than 90\%. When the final percentage of the dying population is more than 90\%, then the corresponding tipping point $t_{0.9}$
becomes infinite, i.e. it is certain that all nodes will die. This case appears only for $f=1$ and large values of $\beta$.

The values of $t_{0.9}$ are 0 for small values of $\beta$, practically independently of the values of $f$. Other than that, the influence of the infection
probability $\beta$ is much weaker than the mortality rate $f$. Large values of $f$ result in small values for $t_{0.9}$, i.e. there is very little time
until a node can feel safe. At small values of $f$, though, this time becomes orders of magnitude larger, and it may take more than 200 steps until the
survival probability reaches 90\%. The results are rather similar among the studied topologies.

When we consider $\Phi(p)$, instead, the topology has a greater influence on the results. For the square lattice, the point $p_{0.9}$ is close to $p_{0.9}=1$
when both $\beta$ and $f$ are large. When $f$ is small this value is closer to $p_{0.9}=0.5$, and vanishes for small $\beta$. In self-organized networks, $p_{0.9}=1$
for large $\beta$ values but now $f$ is small. When $f$ is larger, then this value drops to $p_{0.9}\sim 0.8$. When $\beta$ is small, the influence of $f$ is
negligible, but the values of $p_{0.9}\sim 0.25$ are significantly higher than in the case of lattices ($p_{0.9}\sim 0$).

\subsection*{Functional impact to structure from disease-induced mortality}

Having established how epidemics of this type function over the network structures that we study, we turn our attention into the real damage done by the infection.
The end fraction of dead individuals is an indication of what part of the social structure has survived, but the most important quantity for functional purposes
is the connectivity of the remaining network. Typically, this is described through the size of the largest remaining cluster, $S_{\rm max}$. However, as mentioned
in the Methods section, it is possible that the remaining cluster is connected but the distances are so large that communication in the network may no longer function properly.

Compared to other robustness measures, such as e.g. the largest cluster size or the unique robustness measure, the use of the stability index offers two main advantages,:
a) it can measure the structural damage even if the disease has not eliminated the largest cluster (integral calculated up to $p_{\rm max}$),
and b) the extent of damage can reflect our tolerance for the spatial expansion of communication lengths.

\begin{figure}[h]
\centerline{\resizebox{15cm}{!} { \includegraphics{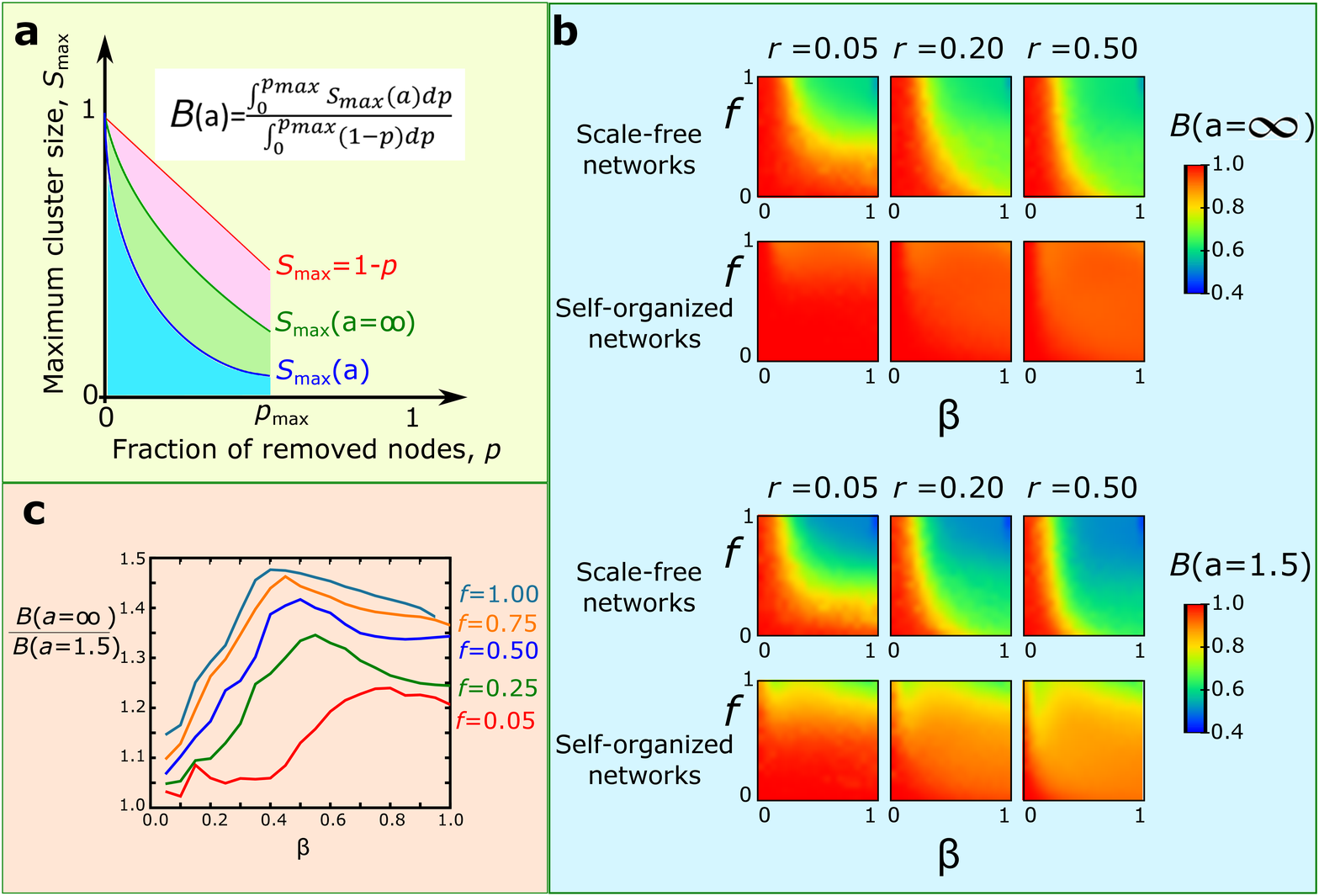}}}
\caption{{\bf Structural impact of the epidemic process.}
(a) Calculation of the quantity $B(a)$. For a given value of $a$ we calculate the ratio of the area under the $S_{\rm max}$ curve over the area under the case
of least possible damage $1-p$. (b) The values of $B(a)$, calculated in scale-free and self-organized networks, as a function of $\beta$ and $f$. The top two
rows correspond to $a=\infty$, and the bottom two rows to $a=1.5$. (c) Comparison of the structural damage measured through the largest cluster size ($a=\infty$)
vs the damage measured through limited-path-percolation ($a=1.5$). The ratio indicates that the largest cluster may overestimate the robustness of the
social structure by a factor of up to $1.5$.
}
\label{fig6}
\end{figure}

In scale-free networks (Fig.~\ref{fig6}b), an increase of either the infection probability or the mortality rate leads to a rapid decrease of the stability
index from $B(\infty)\sim 1$ to $B(\infty)\sim 0.5$. In self-organized networks, though, the variation of the infection probability does not have the same
impact as the mortality rate. These networks are found to be more robust than the random scale-free networks, even though under the same conditions they lose
more nodes due to disease-induced deaths (Fig.~\ref{fig2}). We repeated the same set of simulations for the case of $a=1.5$, i.e. two nodes are not considered
connected if their distance exceeds 1.5 times their distance in the original network. In this case (Fig.~\ref{fig6}b, bottom) the dependence on $\beta$ and $f$
remained qualitatively the same, but now the values of the stability index were considerably lower. 

We quantified this significant variation of the stability index, $B(a)$, with decreasing $a$ (Fig.~\ref{fig6}c) by comparing the ratio of the stability index
$B(\infty)/B(a=1.5)$ for pairs of ($\beta$, $f$). The differences are relatively small for small infection probabilities $\beta$, but they increase as we increase
$\beta$ and become more prominent as we increase $f$. In the most extreme case, the ratio has a value of 1.5, which means that we over-estimate the probability
of the network remaining connected by a factor of 50\%. This can be crucial in border-line cases, where the existence of a connecting cluster would signal the
survival of a group, but in practice the extended damage leads to long connectivity paths that render communication difficult or impossible.
The curves reach a maximum value (i.e. maximum error) for values of $\beta$ in the range of 0.4-0.5, indicating that the error diminishes for higher
infection probabilities, where typically epidemics spread over the entire population.

\section{Discussion}

Disease-induced mortality models can lead to a rapid extinction of the underlying population, but the conditions required for this may be far from trivial.
In particular, a scale-free network topology may accelerate spreading but it also limits the extent of the area that is susceptible to infection.
These conflicting factors can be traced to the effect of the hubs, which can easily reach different parts of the network. However, if a hub remains immune
or removed because of the disease it facilitates disease isolation and communication between different network areas becomes much more difficult.
Contrary to intuition, extensive damage in scale-free networks occurred only for very high probabilities of infection and mortality. When the hubs
become extremely dominant, such as in self-organized networks, then the dominant parameter is the infection probability rather than the mortality rate.
Even small mortality rates can lead to network destruction, as long as the infection probability remains high and preserves the infection in the system. 

We see, then, that the impact of the hubs is not as straightforward as intuition may suggest from their role of bringing all network nodes closer to each other.
All the nodes in lattices are equivalent, but the fraction of removed nodes is systematically higher than in scale-free networks, under the same simulation conditions.
The epidemics duration in lattices is, of course, much longer because of their large-world character. Despite these differences, the survival probability is comparable
in both cases, and a node can feel `safe' from removal after surviving roughly 10 steps.

The fate of the infection spreading depends on the interplay between the model parameters $r$, $f$, and $\beta$ and the structure itself. In general,
small protection loss rates, $r$, protect nodes by providing temporary immunity and possibly allowing the infection to be removed from their neighborhood.
However, this change is mainly quantitative, while qualitatively the behavior remains similar as we increase $r$. Unsurprisingly, the main drivers of
the epidemics are $\beta$ and $f$. Obviously, when both the infection probability and the mortality rate are high, the infection quickly eliminates
the majority of the system. Critically, when only one of these parameters is large or if their values are relatively high, though, then the behavior
largely depends on the structure. For example, a small infection probability in lattices prevents extensive mortality even if $f$ is large.
In self-organized networks we find the opposite picture, where $\beta$ does not influence the outcome but a large mortality rate leads to extended node removal.
These outcomes reveal a much more complicated and nuanced dynamic for the spread of infectious diseases in network-structured populations than have previously been explored.
This suggests that many of the results in the literature, which have been assumed to apply generally to a diversity of diseases and a range of qualitatively similar
network structures, may actually apply only more narrowly to certain ranges of combinations of those descriptors. 

We also found that if we quantify network robustness based on the size of the largest cluster only, we may over-estimate the efficiency of the network by a factor of 50\%.
The natural conclusion is that survival of a connected structure does not necessarily mean that the functionality remains intact, and depending on the communication
requirements the network may have already stopped functioning as intended. To address this, we introduced the stability index suitable for describing the extent of
structural damage during a spreading process. The index can quantify the efficiency of communication in the resulting disease-affected structure by going beyond
the existence of the connected cluster, and taking into account the increase in path lengths. This index incorporates information both from the path lengths and the
dynamics of the spread of the disease. As such, it offers many distinct advantages: a) we do not assess the damage by a binary measure, i.e. the existence or not of
a spanning cluster, b) the stability index can be readily compared across different networks, since it considers the damage up to the point where the infection dies out,
c) the index takes into account the removal history, so that damage done at earlier stages leads to smaller indexes, and d) the structural damage is evaluated according
to the loss of paths, so that differences in clusters with the same number of nodes are still captured by the index.

This study demonstrates the effect of disease-induced mortality in a population, assuming it undergoes one epidemic outbreak that leaves the network weakened,
compared to its initial state. Subsequent outbreaks can accumulate additional damage on the network robustness, but now spreading starts in a different initial structure.
This results to a huge amount of possible combinations (the second disease may have different features than the first). The susceptibility of the resulting networks can be
indirectly found by using the results presented above to apply the same process to the damaged network, instead of the unperturbed structure. We plan to study the effect
of repeated epidemics in a future work.

These results imply that very specific scenarios may offer greater protection from outbreaks that could otherwise compromise populations. This conclusion may be of special
concern in the context of (re)emerging zoonotic infections where populations from multiple host species may be affected in different ways due to differences in physiological
responses to infection. Patterns in the species-to-species paths by which zoonotic diseases reach human populations are dependent on the survival of infected animal populations
at levels that permit continued circulation of disease for long enough to interact with humans (or at least other intermediate animal hosts) to enable transmission.
Such patterns are critical to the metapopulation dynamics in the ecology of infectious diseases \cite{Lloyd,Suzan}, and the models here presented provide greater insights into the driving
forces that may produce these patterns in ways that have gone as yet unexplored. In this way, we may provide an otherwise-missing element needed to estimate zoonotic risks based
on the interaction of epidemiology and social behavior in the reservoir species involved.

Our investigations show nontrivial interactions among the parameters of transmission and mortality risks and the network structure in a more nuanced way than is usually described
when studying disease spread on networks. Additionally, the stability metric presented extends our ability to quantify the expected practical outcomes to populations experiencing
outbreaks beyond the traditional measures to include plausibility of communication, rather than just possibility of communication. Together, these types of analyses over both
epidemiological and topological variations increase our understanding of the extent and timing of population vulnerability to outbreaks of infectious diseases.

\begin{acknowledgments}
We thank the Dept. of Homeland Security for funds in support of this research through the
CCICADA Center at Rutgers, and NSF EaSM grant No. 1049088.
\end{acknowledgments}

\end{document}